\documentclass[aps,preprint,superscriptaddress]{revtex4-1}
\usepackage{lipsum}
\usepackage{epsfig}
\usepackage{amssymb,amsmath,amsthm}
\usepackage{color}
\usepackage[colorlinks=true,linkcolor=blue,citecolor=blue]{hyperref}

\newcommand{\p}{\partial}
\newcommand{\bcdot}{\boldsymbol\cdot}
\newcommand{\bnabla}{\boldsymbol\nabla}

\begin{document}


\title{Pedal underwater motion triggers highly-peaked resonance on water waves}




\author{\copyright Juan F. Mar\'in}
\email[]{juan.marin.m@usach.cl}
\affiliation{Departamento de F\'isica, Universidad de Santiago de Chile, Usach, Av. V\'ictor Jara 3493, Estaci\'on Central, Santiago, Chile.}

\author{Alexander Egli}
\affiliation{Departamento de F\'isica, Universidad de Santiago de Chile, Usach, Av. V\'ictor Jara 3493, Estaci\'on Central, Santiago, Chile.}

\author{Isis Vivanco}
\affiliation{Departamento de F\'isica, Universidad de Santiago de Chile, Usach, Av. V\'ictor Jara 3493, Estaci\'on Central, Santiago, Chile.}

\author{Bruce Cartwright}
\altaffiliation{Also at Pacific Engineering Systems International, 277\textendash 279
Broadway, Glebe, New South Wales 2037, Australia}
\affiliation{The University of Newcastle, Callaghan, New South Wales 2308,
Australia}

\author{Leonardo Gordillo}
\affiliation{Departamento de F\'isica, Universidad de Santiago de Chile, Usach, Av. V\'ictor Jara 3493, Estaci\'on Central, Santiago, Chile.}


\date{\today}

\begin{abstract}
Pedal wavemakers that generate surface gravity waves through bed orbital motion have been shown to produce particle-excursion patterns that mimic deep-water wave behaviour but in finite-depth channels. In this article, we report that gravity waves in a general viscous fluid can resonate through the action of pedal wavemakers. We analyse the linear response of waves in an infinite channel in terms of the displacement amplitude, frequency, and wavelength of the bottom action. We show that the system behaves as a long-pass filter in space and a high-pass filter in time with a sharp resonance affected by viscosity. Furthermore, we propose a protocol to design deep gravity waves with an on-demand wavelength in a finite-depth water channel. Our theoretical framework agrees with numerical simulations using Smoothed Particle Hydrodynamics. Our results thus quantify the performance of pedal wavemakers and provide essential formulas for industrial and computational applications of the pedal wavemaking technique, useful both in hydraulics and coastal engineering problems. 
\end{abstract}


\maketitle



\section{\label{sec:Introduction}Introduction}

Some of the most intriguing wave phenomena occurring at the surface of the deep ocean lie within the long-wavelength spectrum \citep{Massel2013, Munk1950}. One example is the complex dynamics observed at the sea surface after storms and hurricanes \citep{Marsooli2018}, after atmospheric forcing and perturbations from persistent high-speed winds create intricate wave patterns and induce complex velocity fields. Another example is tsunamis, which appear due to the sudden motion of a large volume or boundary in contact with the fluid. Short-range tsunamis are commonly generated when the external forcing is applied at the free surface of the fluid, for instance, when large volumes of solid material collapse on the surface of the water due to landslides \citep{Sarlin2021} or glacier ice fragmentation \citep{Wolper2021}. Long-scale tsunamis are, in turn, generated when the forcing is exerted at the seabed, such as in submarine volcano eruptions and earthquakes  \citep{Kajiura1963}. In the long-wavelength limit, determining the optimal transfer of motion from a source to the resulting waves on a fluid remains essential for several marines and coastal engineering problems.

In the laboratory, the artificial generation of long waves in water channels using conventional wavemakers, through paddles and pistons, has been the standard in experimental and numerical fluid dynamics \citep{Dean1991}. However, paddles and pistons do have issues: they generate undesired evanescent waves and their performance at long wavelengths is poor \citep{Dean1991, Havelock1929, Ursell1960, Lighthill1978}. Even in the limit of zero viscosity, conventional wavemakers have to oscillate with an amplitude comparable to the wavelength, generating a wave whose amplitude is a fraction of the wavemaker amplitude. In recent work \citep{Vivanco2021}, a solution for this problem was proposed by introducing a new type of wave generator: \emph{the pedal wavemaker}. Inspired in wave generation by earthquakes and the excursion of fluid particles under waves, given by the Airy theory of inviscid deep-water waves, the technique consists of moving the bottom of a wave channel with a pedalling-like orbital motion, as illustrated in figure~\ref{Fig:01}. Remarkably, such generally elliptical orbital motion can also be imposed in slightly viscous fluids and particle excursion is affected by the development of oscillatory Stokes boundary layers in the bottom and surface\citep{Vivanco2021}. All these features suggest promising applications in studies of fluid-structure interactions in hydraulics, including numerical simulations and laboratory-scale tests of oil platforms and ships under extreme marine conditions \citep{Cartwright2010, Cartwright2012, Groenenboom2019, Groenenboom2021}.

The efficiency of pedal wavemaking is can be analysed in terms of the system response to different wavelengths and frequencies. Theoretical analysis plays also a crucial role in quantifying dissipation and momentum transfer between the fluid's bottom and surface. Such a study can set the basis for a protocol for efficiently generating long gravity waves in artificial water channels under laboratory conditions, emulating deep-water conditions using finite-depth water tanks.

This article studies the linear response of surface gravity waves generated by pedal wavemaking in a viscous fluid. By analysing the hydrodynamic equations for viscous fluids under appropriate boundary conditions, we show by analytical methods that it is possible to observe resonance, wave synchronisation, and filtering-like behaviour in the system as a function of the wavelength, frequency, and viscosity. We demonstrate that under resonant conditions, a small-amplitude motion of the pedal wavemakers sets large-amplitude waves at the surface. Our theoretical findings agree with numerical simulations using the Smoothed Particle Hydrodynamics (SPH) method. Our results are organised as follows: In \S \ref{Sec:Resonance}, we introduce the theoretical model and study the linear response of the free surface of the fluid driven by pedal wavemakers at the bottom. \S \ref{Sec:Design} shows how to design long gravity waves emulating deep-water behaviour. The SPH numerical methodology is briefly outlined in \S \ref{Sec:SPH}, where we compare the outcome from numerical simulations with the expected theoretical results. We conclude with our final remarks in \S \ref{Sec:Conclusions}.

\section{Resonance of long gravity waves
\label{Sec:Resonance}}

\begin{figure}
    \centering
    \scalebox{1}{\includegraphics{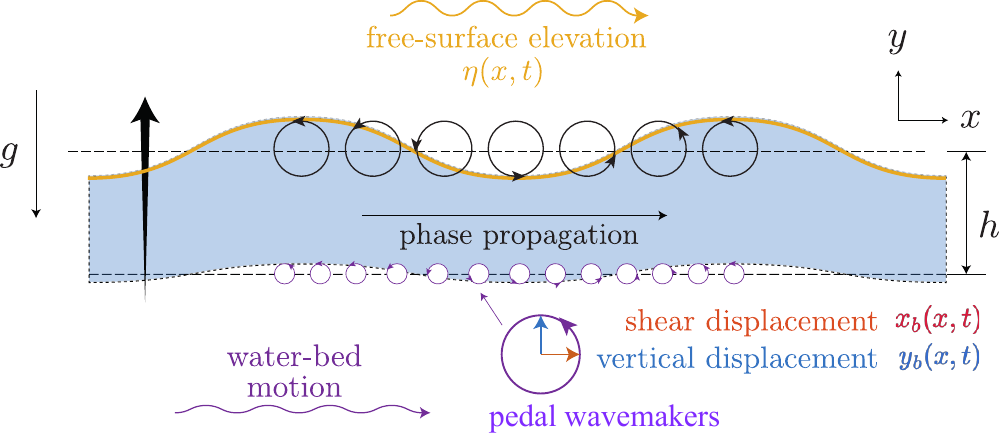}}
    \caption{Pedalling wavemaking: a two-dimensional infinite fluid domain placed over a soft bottom whose motion is prescribed by pedal wavemakers. The resulting surface wave has a phase propagation towards the positive $x$-direction.}
    \label{Fig:01}
\end{figure}

Consider the system shown in figure~\ref{Fig:01}, where gravity waves are generated at the surface of a viscous fluid using pedal wavemaking \citep{Vivanco2021}. The domain is an infinite two-dimensional fluid of depth $h$ placed over a soft deformable bottom. The pedalling-like motion of the wavemakers at $y=-h$ is set to exert a travelling wave forcing. The equations describing the system are the Navier-Stokes (NS) equations for incompressible fluids,
\begin{subequations}
\label{eq:Ns-incompress}
\begin{equation}
\p_{t}\mathbf{u}+(\mathbf{u}\bcdot\bnabla)\mathbf{u}=-\frac{1}{\rho}\bnabla P+\nu\nabla^{2}\mathbf{u}+\mathbf{g}\label{eq:NS}
\end{equation}
\begin{equation}
\bnabla\bcdot\mathbf{u}=0,\label{eq:incompress}
\end{equation}
\end{subequations}
where $\mathbf{u}=u\mathbf{i}+v\mathbf{j}$ is the velocity field, $P$ is the pressure, $\nu$ is the kinematic viscosity, $\mathbf{g}=-g\mathbf{j}$ is the acceleration of gravity, and $\rho$ is the density of the fluid (water). 

The water-bed motion induced by pedal wavemakers has a phase propagation in the positive $x$-direction, with prescribed wavenumber $k$ and angular frequency $\omega$. Such motion can be decomposed in vertical and horizontal periodic displacements, which can be described by the complex fields $y_b:\mathbb{R}^2\to\mathbb{C}$ and $x_b:\mathbb{R}^2\to\mathbb{C}$, respectively, and are given by
\begin{subequations}
\label{eq:input}
\begin{equation}
x_b(x,t)=\mathrm{i}\chi\exp[\mathrm{i}\left(kx-\omega t\right)],\label{Eq:InputHorizontal}
\end{equation}
\begin{equation}
y_b(x,t)=\zeta\exp[\mathrm{i}\left(kx-\omega t\right)],\label{Eq:InputVertical}
\end{equation}
\end{subequations}
where $\chi$ and $\zeta$ are the amplitudes of the pedalling motion. The input drive of the system is given by \eqref{eq:input}, satisfying the no-slip boundary condition at the bottom,
\begin{subequations}
	\label{eq:bottomboundary}
	\begin{equation}
		\label{eq:noslip1}
		\left.\p_tx_b-u\,\right|_{y=-h}=0,
	\end{equation}
		\begin{equation}
		\label{eq:noslip2}
		\left.\p_ty_b-v\,\right|_{y=-h} =0.
	\end{equation}
\end{subequations}
%
The linear response under study, i.e. the output of the system \eqref{eq:Ns-incompress}, to the input drive \eqref{eq:input} is a surface gravity wave, denoted as $\eta:\mathbb{R}^2\to\mathbb{C}$, which in the linear regime has the same frequency $\omega$ and wavenumber $k$ as the bottom motion,
\begin{equation}
\eta(x,t)=\eta_0\exp[\mathrm{i}\left(kx-\omega t\right)],\label{Eq:Output}
\end{equation}
where $\eta_0$ is a complex amplitude. Such surface wave must comply with the linear boundary conditions at the free surface, namely
\begin{subequations}
	  	\label{eq:boundary}
	\begin{equation}
		\label{eq:kinematic}
		\left.\p_t\eta-v\,\right|_{y=\eta}=0,
	\end{equation}
	\begin{equation}
		\label{eq:normal_stress}
		\left.2\nu\p_{y}v-\frac{P}{\rho}\,\right|_{y=\eta}=0,
	\end{equation}
	\begin{equation}
		\label{eq:tangential_stress}
		\left.\p_{x}v+\p_{y}u\,\right|_{y=\eta}=0.
	\end{equation}
\end{subequations}
Equation~\eqref{eq:kinematic} is the kinematic condition, whereas \eqref{eq:normal_stress} and \eqref{eq:tangential_stress} are the conditions for the normal and tangential stress, respectively. 

\citet{Vivanco2021} solved the linearized \eqref{eq:Ns-incompress} using the Helmholtz decomposition \citep{Lamb1932}, $\ensuremath{\mathbf{u}=\bnabla\phi+\bnabla\times\left(\psi\mathbf{k}\right)}$, where $\phi$ is the velocity potential and $\psi$, the Stokes stream function. From the linear NS equations, one obtains that the velocity potential $\phi$ is
\begin{subequations}
\label{Eq:Potential}
    \begin{equation}
	\label{eq:incompress2}
	    \nabla^2\phi =0,
	\end{equation}
    \begin{equation}
		\label{eq:potential}
		\p_t\phi+\frac{P}{\rho}+gy=0,
	\end{equation}
\end{subequations}
whereas the stream function $\psi$ is governed by
\begin{equation}
    \label{eq:stream}
	\nu\nabla^2\psi-\p_t\psi=0.
\end{equation}

\begin{figure}
\begin{center}
\scalebox{0.2}{\includegraphics{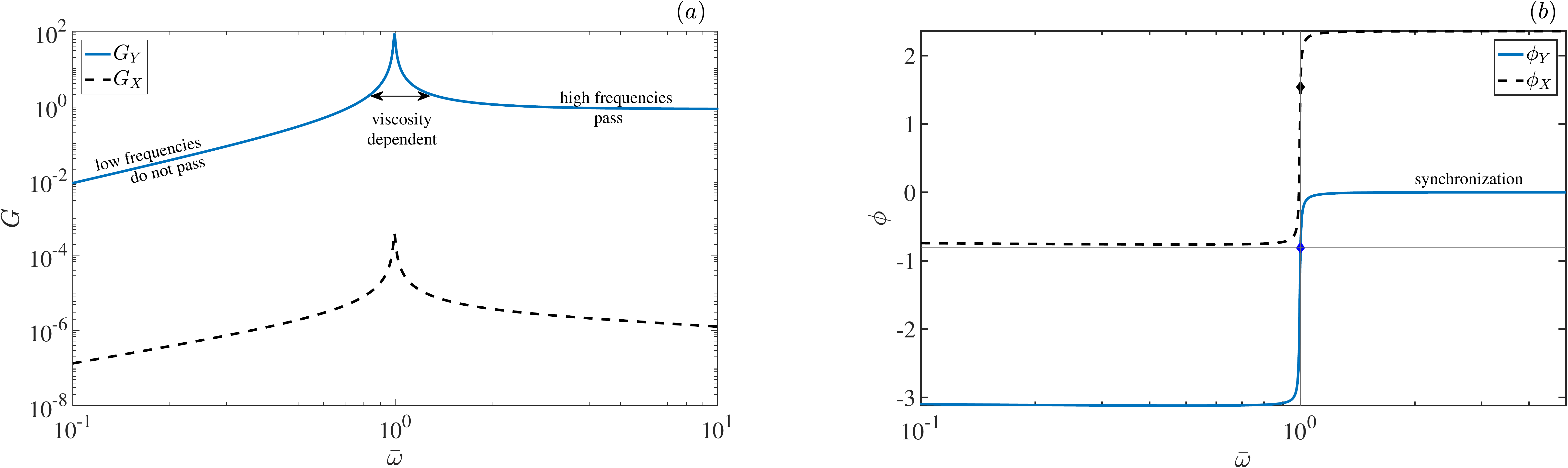}}\\
\scalebox{0.2}{\includegraphics{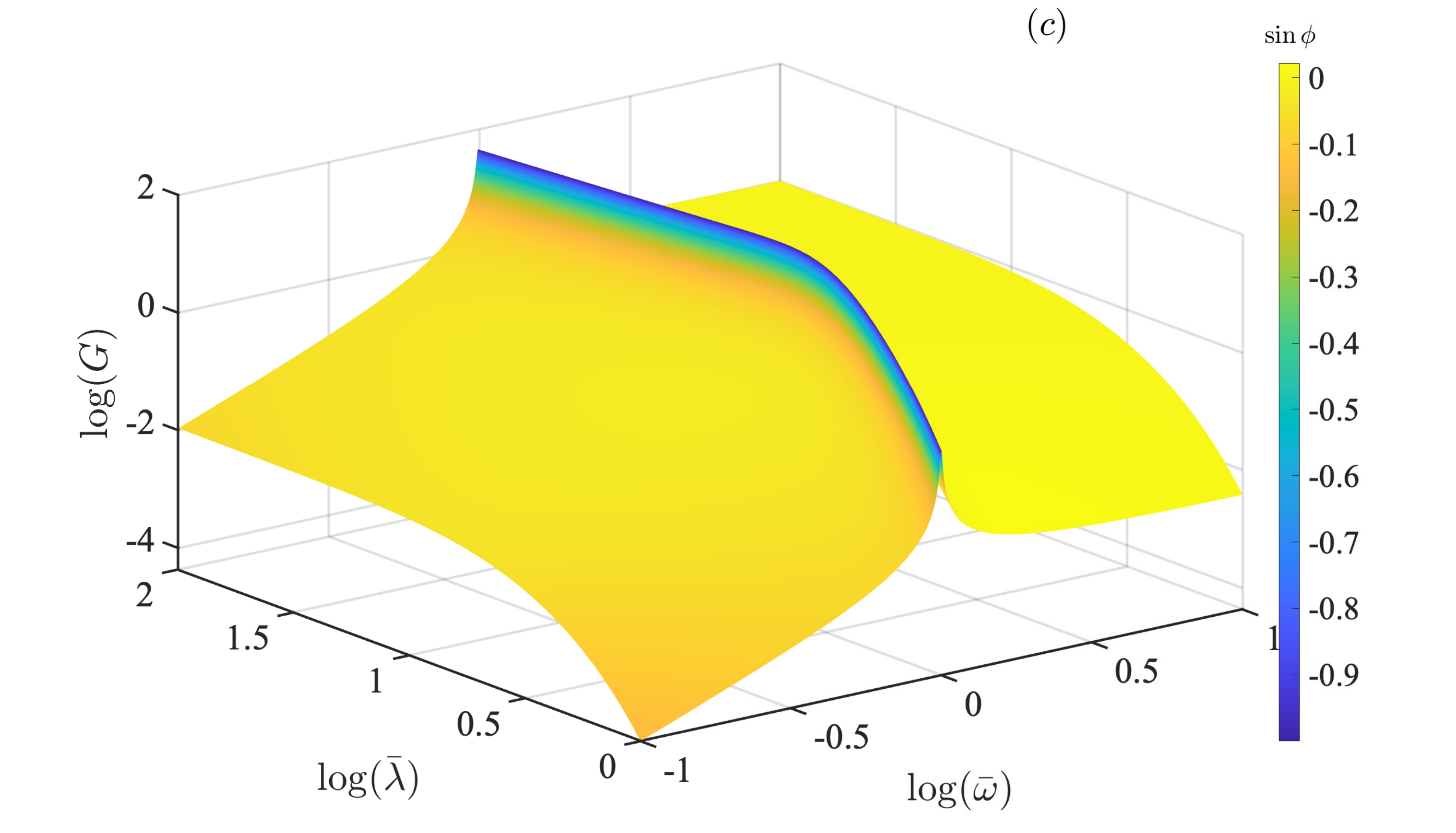}}
\end{center}
 \caption{Typical frequency-response curves for a $40\,m$-deep channel filled with a fluid of viscosity $\nu=1800$ cSt ($\Theta=4.55\times10^{-4}$). \textbf{\textit{(\textit{a})}}
 The gain of the vertical and horizontal components (denoted as $G_Y$ and $G_X$, respectively) of the pedal-wavemaking drive for $\bar\lambda=10$, with a resonant peak near the frequency $\omega_0=\omega_{\infty}\sqrt{\tanh{kh}}$. The system behaves as a filter of low frequencies. \textbf{(\textit{b})} The phases $\phi_X=\arg(X)$ and $\phi_Y=\arg(Y)$, revealing
 synchronisation at frequency ranges slightly above the resonance condition (indicated by diamonds). \textbf{(\textit{c})} Gain as a function of the wavelength and the frequency of the pedal-wavemaking force. The sine of the phase is shown in the colour scale. \label{Fig:Resonance}}
\end{figure}

Hereon, we use the dimensionless wavenumber $\bar{k}=kh$ and frequency $\bar{\omega}=\omega/\omega_0$, where $\omega_0=\sqrt{\tanh\bar{k}}$ is the dispersion relation of gravity waves normalised by $\omega_{\infty}=\sqrt{gk}$. Likewise, the amplitude of the surface wave, the horizontal bottom motion, the vertical bottom motion, the vertical coordinate, and horizontal coordinate in dimensionless form are $\bar\eta_0=\eta_0/h$,  $\bar\chi=\chi/h$,  $\bar\zeta=\zeta/h$, $\bar{y}=y/h$, and $\bar{x}=x/h$, respectively.

We consider the exact solutions for the velocity potential and the stream function \citep{Vivanco2021}. After substitution into the kinematic boundary condition \eqref{eq:kinematic}, the complex amplitude at the surface is $\bar\eta_{0}=X\bar\chi+Y\bar\zeta$, where
\begin{subequations}
\label{Eq:Theoretical}
\begin{equation}
\label{XAmplitude}
 X(\bar\omega, \bar k):=\frac{\mathrm{i}\bar\Theta}{\Xi(\bar\omega, \bar k)}\left(\frac{1}{\alpha}\sinh\alpha \bar{k}-\sinh \bar{k}\right),
\end{equation}
\begin{equation}
\label{YAmplitude}
 Y(\bar\omega, \bar k):=\frac{\left(1+\mathrm{i}\bar\Theta\right)\cosh\alpha\bar{k}-\mathrm{i}\bar\Theta\cosh\bar{k}}{\Xi(\bar\omega, \bar k)}.
\end{equation}
\end{subequations}
The function $\Xi$ is given by

\begin{eqnarray}
\Xi(\bar\omega, \bar k)&=&\cosh\alpha\bar{k}\cosh\bar{k}\left(1-\frac{1}{\bar\omega^{2}}\right)\left(1-\frac{\mbox{tanhc}\,\alpha \bar{k}}{\mbox{tanhc}\,\bar{k}}\right)+\frac{\mbox{sinhc}\,\alpha \bar{k}}{\mbox{sinhc}\,\bar{k}} \nonumber\\
&+&\bar{k}^{2}\left[\mbox{sinhc}^2\left(\frac{\bar{k}(\alpha +1)}{2}\right)+\mbox{sinhc}^2\left(\frac{\bar{k}(\alpha -1)}{2}\right)\right].
\end{eqnarray}
%
Here, $\alpha^{2} =1-2\mathrm{i}\bar\omega/\bar\Theta\sqrt{\bar k^3}$ is a dimensionless quantity and $\bar\Theta =2\nu/\sqrt{gh^3}$ is the dimensionless viscosity of the fluid. The functions $\mbox{sinhc}(x)=(\sinh x)/x$ and $\mbox{tanhc}(x)=(\tanh x)/x$ are the hyperbolic sine cardinal \citep{Sanchez2012} and hyperbolic tangent cardinal, respectively. 

The total gain in the system is $G=\sqrt{|X|^2+|Y|^2}$, where $X$ and $Y$ are given by \eqref{XAmplitude} and \eqref{YAmplitude}, respectively as the gain has contributions from tangential and normal components of the pedal-wavemaking motion at the bottom. Figure~\ref{Fig:Resonance} shows typical response curves of the system as a function of the normalised frequency and wavelength for the given values of parameters. Figure~\ref{Fig:Resonance}(\textit{a}) evidences that $|X|\ll |Y|$, which means that the contribution of the horizontal component of the bottom motion to the total gain is negligible.

\begin{figure}
\begin{center}
\scalebox{0.36}{\includegraphics{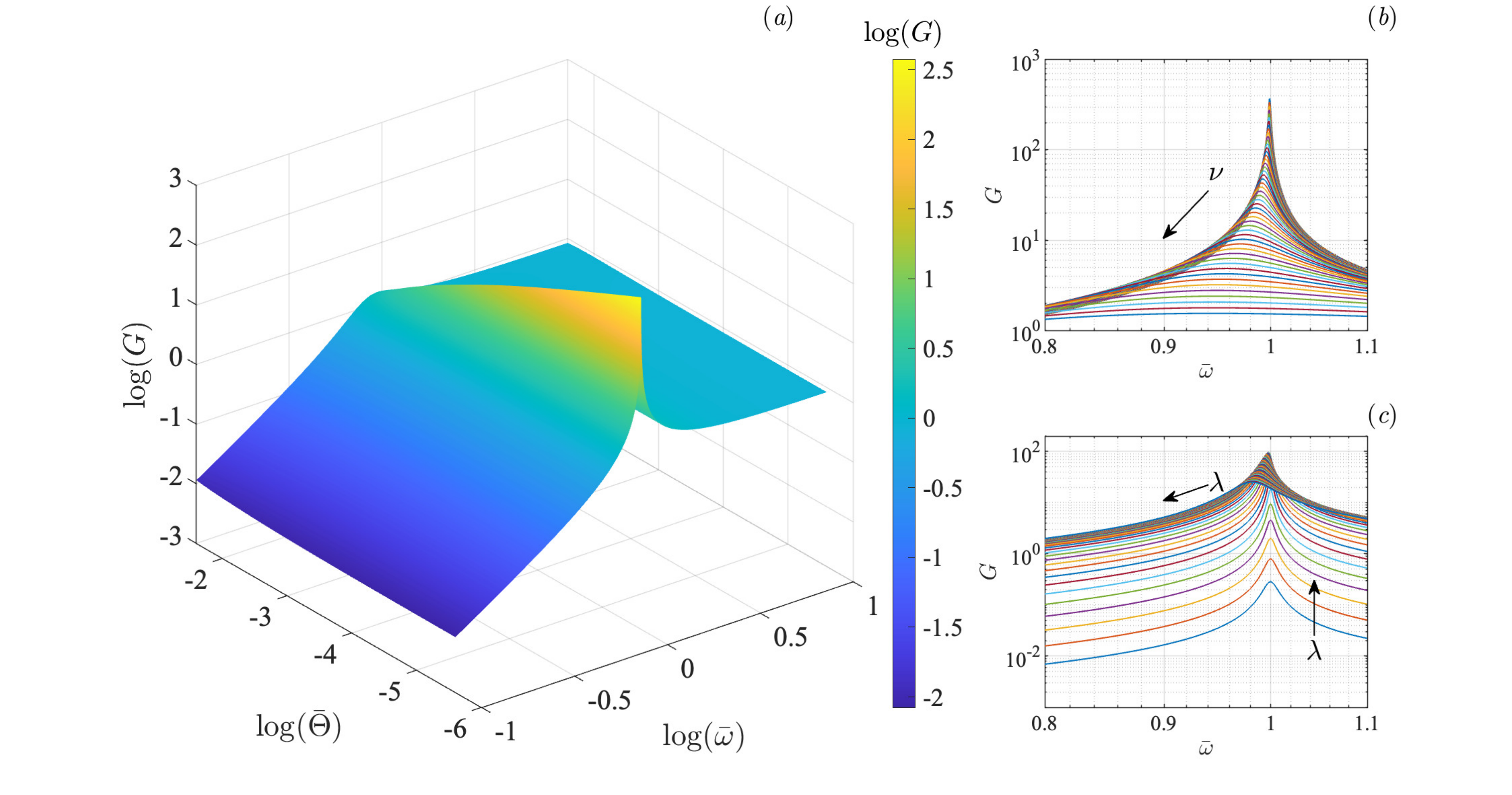}}\\
\scalebox{0.4}{\includegraphics{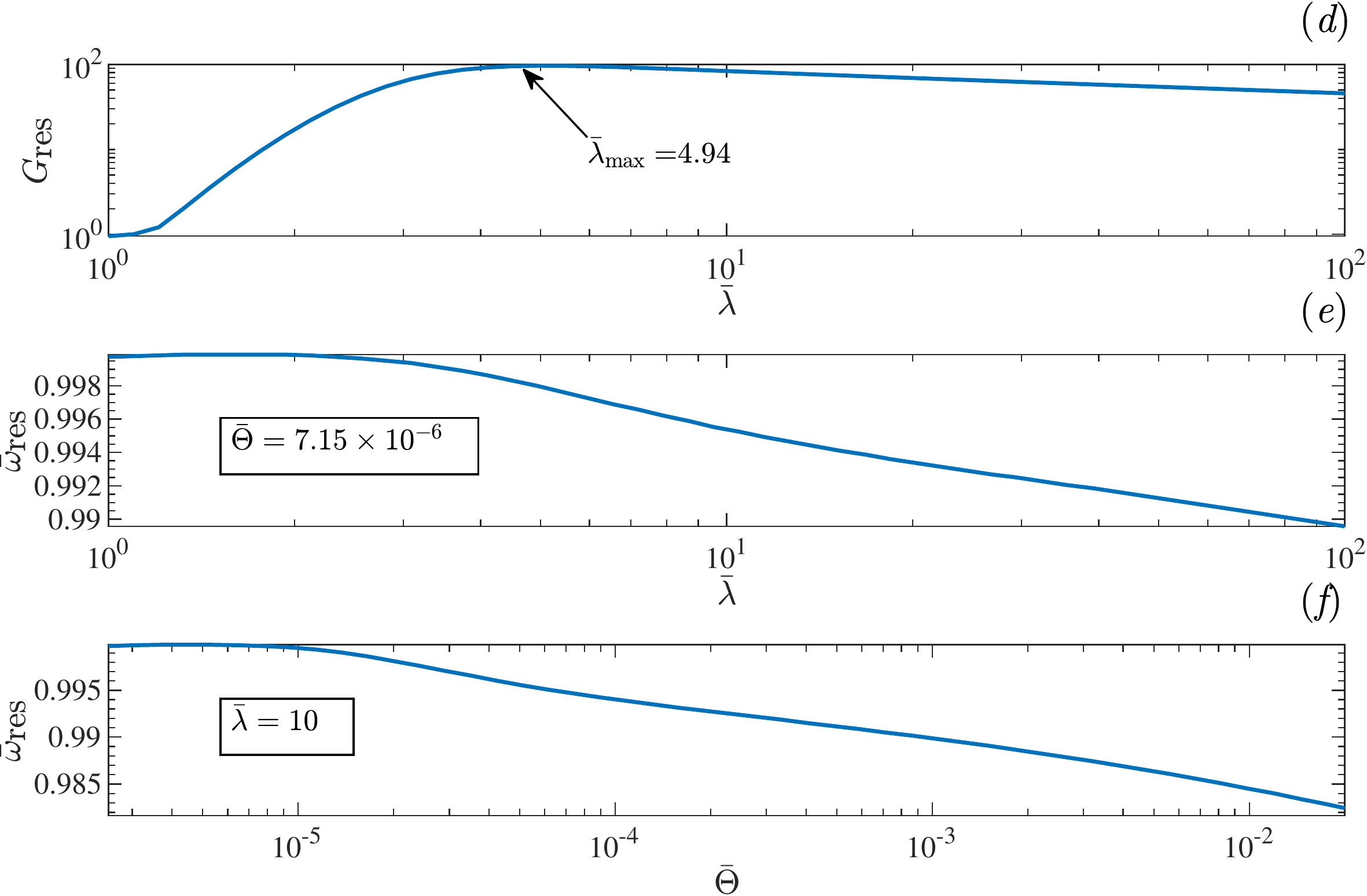}}
\end{center}
 \caption{Dependence of the resonant behaviour on the viscosity and wavelength. \textbf{(\textit{a})} Gain of the system as a function of the dimensionless frequency $\bar\omega$ and viscosity $\bar\Theta$ of the fluid for $\bar\lambda=10$. \textbf{(\textit{b})} Gain as a function of the frequency for different values of the viscosity and fixed wavelength $\bar\lambda=10$. The arrow indicates the behaviour as the viscosity increases. \textbf{(\textit{c})} Gain as a function of the frequency for different values of the wavelength and fixed viscosity $\bar\Theta=4.5\times10^{-4}$. The arrows indicate the behaviour as the wavelength $\lambda$ increases. \textbf{(\textit{d})} The maximum gain in the system $G_{\max}$ as a function of the wavelength for $\bar\Theta=7.15\times10^{-6}$. The wavelength for maximum gain is indicated with an arrow. \textbf{(\textit{e})} Frequency of the resonant mode as a function of the wavelength of the gravity wave and \textbf{(\textit{f})} the viscosity of the fluid. \label{Fig:Resonance_Viscosity}}
\end{figure}

Figure~\ref{Fig:Resonance}(\textit{a}) shows that the total gain $G$ of the system exhibits a resonance peak for $\bar\omega=\bar\omega_{\footnotesize\mbox{res}}\simeq1$, i.e, for $\omega\simeq\omega_0$. At such precise frequency, the motion is optimally transferred from the pedal wavemakers to the surface of the fluid, generating a wave that mimics deep-water behaviour if $\lambda\simeq 2\pi h$ ($\omega_0\simeq\omega_{\infty}$). Such resonance feature further supports the pedal-wavemaking technique in recreating deep long-wavelength waves in a finite water channel \citep{Vivanco2021}. Figure~\ref{Fig:Resonance}(\textit{a}) depicts a strong and persistent decay of the gain as the frequency decreases below  $\bar\omega_{\footnotesize\mbox{res}}$. In other words, the system behaves like a filter for low frequencies. In the other limit of the spectrum, the gain slowly decays until it reaches a plateau with unit gain, which means that the system behaves as a high-pass filter. Similar behaviour has been reported in impulsively forced water waves in the context of tsunami generation \citep{Kajiura1963, Hammack1973, Jamin2015}.

As the frequency becomes faster than $\bar\omega_{\footnotesize\mbox{res}}$, generated waves are gradually synchronised with the bottom motion. This is evidenced in the phase-response curves shown in figure~\ref{Fig:Resonance}(\textit{b}), which show the phases $\phi_X=\arg(X)$ and $\phi_Y=\arg(Y)$ of the gravity wave as a function of the input frequency $\bar\omega$. Indeed, we observe that $\phi_Y\to0$ as $\bar\omega$ increases above $\bar\omega_{\footnotesize\mbox{res}}$, and the system synchronises in a broad bandwidth of frequencies above the resonant condition. The system synchronisation above resonance means that the surface of the fluid and the whole fluid layer perfectly follows the bottom vertical motion. Notice that the synchronising effect of $\phi_X$ is negligible. 

Figure~\ref{Fig:Resonance}(\textit{c}) summarises the system response in terms of the gain as a function of both the frequency and wavelength of the pedal-wavemaking motion. We observe that the high-pass filtering behaviour is preserved and that a resonant condition exists for every wavelength of the forcing. An important consequence is that if the prescribed motion of the pedal wavemakers is given by a random distribution of wavelengths and frequencies, only those components with high frequencies and long wavelengths will contribute to the formation of surface waves. This property may find relevant applications in the recreation of oceanic waves in an artificial wave tank since long waves are generated in the ocean from storms, earthquakes, and gravitational tides \citep{Munk1950}.

Figure~\ref{Fig:Resonance}(\textit{c}) also evidences that the pedal wavemakers are inefficient in generating short-wavelength waves. For $\bar\lambda\simeq1$ we found $G\simeq0.1$, which means that the amplitude of the gravity wave is only a 10\% of the amplitude of the bottom motion at the given viscosity. However, the resonant peak monotonically increases with the wavelength. The gain reaches a saturation value around $\bar\lambda\simeq10$, i.e. $G$ remains almost constant for $\bar\lambda>10$, as evidenced in figure~\ref{Fig:Resonance}(\textit{c}). Remarkably, the gain at resonance is large for long wavelength waves if the viscosity is small. Indeed, in figure~\ref{Fig:Resonance}(\textit{c}) the amplitude of the generated waves is two orders of magnitude larger than the amplitude of the bottom motion. The calculations were performed for a $40\,\mbox{m}$-deep channel and show a large resonance, even though the fluid is 1800 times more viscous than water.

Figure~\ref{Fig:Resonance_Viscosity} depicts the effect of the viscosity on the resonance condition. In figure~\ref{Fig:Resonance_Viscosity}(\textit{a}) we observe that the gain at resonance decreases as the viscosity increases, as expected. Moreover, we observe that the resonance peak widens as the fluid becomes more viscous. We show this in more detail in figure~\ref{Fig:Resonance_Viscosity}(\textit{b}). The increase of viscosity also shifts the resonant frequency to lower values.

In figure~\ref{Fig:Resonance_Viscosity}(\textit{c}) we show different resonant curves obtained at different wavelengths for $\bar\Theta=4.5\times10^{-4}$. The gain is low at small wavelengths, as we had previously observed in figure~\ref{Fig:Resonance}(\textit{c}). However, notice that as the wavelength increases, the resonance curve increases uniformly until it reaches some optimal value $\bar\lambda_{\max}$. A further increase of the wavelength above $\bar\lambda_{\max}$ produces a broadening of the resonant peak and a decrease in the maximum gain. We confirm this observation in Fig~\ref{Fig:Resonance_Viscosity}(\textit{d}), where we show the gain at resonance as a function of the wavelength. Indeed, the gain is maximum at the precise wavelength $\bar\lambda_{\max}=4.498$, which is almost five times the depth of the water channel. The maximum gain exhibits a small decay when increasing the wavelength further from this optimal value. Thus, there is also an optimal wavelength that maximises the gain. 

\section{Deep long gravity waves in a finite-depth tank \label{Sec:Design}}

From the results in \S \ref{Sec:Resonance}, we can efficiently induce long gravity waves for a given fluid in a finite-depth tank with an on-demand wavelength. Figure~\ref{Fig:Resonance_Viscosity}(\textit{e}) shows the frequency $\bar\omega_{\footnotesize\mbox{res}}$ at which the pedal wavemakers must oscillate to generate a wave with some arbitrary wavelength for the given value of the viscosity. Notice that $\bar\omega_{\footnotesize\mbox{res}}\sim1$ for the whole interval of wavelengths shown, $\bar\lambda\in[1,\,100]$. This means that the resonant gravity waves oscillate nearly as if they were under deep-water conditions (in the physical variables, with frequency $\omega\sim\omega_{\infty}$). As noticed in \citet{Vivanco2021}, the particle orbits in the bulk of the fluid also have deep-water-like behaviour except for the influence of thin boundary layers at the free surface and the moving bottom. Here, we have also demonstrated that the wave generation technique is tunable because the displacement is optimally amplified for an on-demand wavelength.

We also found that the properties of the designed wave are very robust to viscosity effects. Figure~\ref{Fig:Resonance_Viscosity}(\textit{f}) shows that $\bar\omega_{\footnotesize\mbox{res}}\sim1$ inside a wide range of small viscosity values, $\bar\Theta\in[10^{-6},\, 10^{-2}]$ for an arbitrary wavelength $\bar\lambda=10$. In contrast, for a substantial viscosity, we expect that the wavelength and frequency of the gravity wave will have larger deviations from the deep-water behaviour due to the broadening of the resonance peak, as we previously discussed in figure~\ref{Fig:Resonance_Viscosity}(\textit{b}).


\section{Numerical simulations \label{Sec:SPH}}

\subsection{SPH formulation}

We perform numerical simulations of the hydrodynamic equations using the \emph{Smoothed Particle Hydrodynamics} (SPH) method \citep{Lucy1977, Gingold1977}. In the SPH method, the density, pressure, and velocity field variables and gradients are obtained numerically from a Lagrangian formulation of the NS equations \eqref{eq:Ns-incompress} using a combination of two approximation techniques: the \emph{kernel approximation} and the \emph{particle approximation}. In the following, we give a brief overview of the SPH formulation used in this work based on the Open Software PySPH \citep{PySPH}. A more comprehensive and detailed approach is given by Liu and Liu \citep{Liu2010}, and Monaghan \citep{Monaghan2005, Monaghan2012}.

In the kernel approximation, the continuity and momentum equations are multiplied by a kernel function $W=W(|\mathbf{r}-\mathbf{r}'|;h_w)$ and integrated along $\mathbf{r}'$ in the fluid domain, leading to the so-called weighted integrals. The kernel $W$ is a smooth, normalised, and monotonically decaying function of the distance $|\mathbf{r}-\mathbf{r}'|$, typically with a bell-like shape. Formally, it is assumed that $W(|\mathbf{r}-\mathbf{r}'|;h_w)\to\delta(|\mathbf{r}-\mathbf{r}'|)$ as $h_w\to0$, where $\delta$ is the Dirac delta function. Moreover, kernels are usually chosen to vanish for $|\mathbf{r}-\mathbf{r}'|>h_w$, i.e. they have a \emph{compact support domain} of radius $h_w$ around $\mathbf{r}'$. This first approximation step gives the field variables and gradients as smoothed, local space-averaged values around each fluid particle placed at $\mathbf{r}'$, where only fluid particles close enough contribute to the space averaging. The radius $h_w$ of the compact support is known as the \emph{smoothing length}.

The formal space discretisation of the fluid domain is performed using particle approximation. Weighed integrals are estimated as discrete sums on a set of fluid particles. The resulting discrete SPH formulation for the continuity and momentum equations for the $i$-th particle is
\begin{subequations}
\label{Eq:SPHEquations}
\begin{equation}
\label{Eq:SPHContinuity}
    \frac{\mathrm{d}\rho_i}{\mathrm{d}t}=\sum_{j\neq i}m_j\mathbf{u}_{ij}\bcdot\bnabla_iW_{ij}(h_w),
\end{equation}
\begin{equation}
\label{Eq:SPHMomentum}
    \frac{\mathrm{d}\mathbf{u}_i}{\mathrm{d}t}=-\sum_{j\neq i}m_j\left(\frac{P_i}{\rho_i^2}+\frac{P_j}{\rho_j^2}\right)\bnabla_iW_{ij}(h_w)+\mathbf{\Pi}_i+\mathbf{g}_i,
\end{equation}
\end{subequations}
where $W_{ij}(h_w):=W(\mathbf{r}_{ij};h_w)$, $\mathbf{r}_{ij}:=\mathbf{r}_i-\mathbf{r}_j$, $\mathbf{u}_{ij}:=\mathbf{u}_i-\mathbf{u}_j$, and $\bnabla_i$ denotes a gradient with respect to the coordinates of the $i$-th particle. The artificial viscous acceleration $\Pi_{i}$ is a numerical term originally introduced to control instabilities appearing in problems with shock waves \citep{Monaghan2005, Liu2010, Monaghan2012, PySPH}. We use the SPH formulation of the entropically damped artificial compressibility \citep{Ramachandran2019, PySPH},
\begin{equation}
\label{Eq:ViscositySPH}
\mathbf{\Pi}_{i}:=2\alpha h_w C\sum_{j\neq i}\frac{   m_j}{\rho_i+\rho_j}\,\frac{\mathbf{r}_{ij}\bcdot\bnabla_iW_{ij}(h_w)}{\left(|\mathbf{r}_{ij}|^2+\epsilon h_w^2\right)}\mathbf{u}_{ij},
\end{equation}
where $\alpha=0.2$, $C=442.94\,\mbox{cm/s}$, and $\epsilon=0.01$ is a parameter introduced to avoid singularities when $|\mathbf{r}_{ij}|$ vanishes. The pressure of the $i$-th particle is specified by its density via the usual equation of state for weakly compressible fluids, namely
\begin{equation}
    \label{Eq:EqState}
    P_i=\frac{P_0}{\gamma}\left[\left(\frac{\rho_i}{\rho_0}\right)^{\gamma}-1\right],
\end{equation}
where $P_0$ and $\rho_0$ are the reference pressure and density, respectively. For fluids, it is common to use $\gamma=7$. Finally, time integration of Eqs.~\eqref{Eq:SPHEquations} can be achieved using standard time-integration methods for ordinary differential equations. In this work we used a second order predictor-corrector integrator with $\Delta t=1\times10^{-5}\,\hbox{s}$.

Our numerical setup is schematised in figure~\ref{Fig:04}(\textit{a}). A fluid layer of density $\rho_0=1\,\hbox{gr/cm}^3$ and depth $h=1\,\hbox{cm}$ is initially at rest in a fluid domain of length $L=20\,\hbox{cm}$. We used $N=2343$ fluid particles and $852$ solid particles for the bottom, all with diameter $D=0.1\,\hbox{cm}$. We used a cubic-spline kernel with a smoothing length of $h_w=0.13\,\hbox{cm}$. To avoid the effects of reflections and simulate an infinite large domain, we impose periodic boundary conditions in the $x$-axis. In this pure SPH formulation, the solid bottom is represented by discrete solid particles, as shown in the inset of figure~\ref{Fig:04}(\textit{a}). We use three layers of solid SPH particles for better performance of the bottom boundary and to avoid the percolation of fluid particles during the bottom pedal-wavemaking motion. The boundary condition at the moving bottom is formulated in terms of a generalised wall boundary condition   \citep{Adami2012}, which is based on a local force balance between the bottom solid particles and the fluid particles to prevent wall penetration. We implemented the weakly-compressible SPH formulation   \citep{Hughes2010} for better computational performance in the simulation of the free-surface flow, choosing $P_0$ in \eqref{Eq:EqState} large enough to keep the density fluctuations small ($\sim 1\%$) \citep{Monaghan2005}.

\subsection{Numerical protocol and discussion}

\begin{figure*}
    \centering
    \begin{minipage}{0cm}
    \scalebox{0.5}{\includegraphics{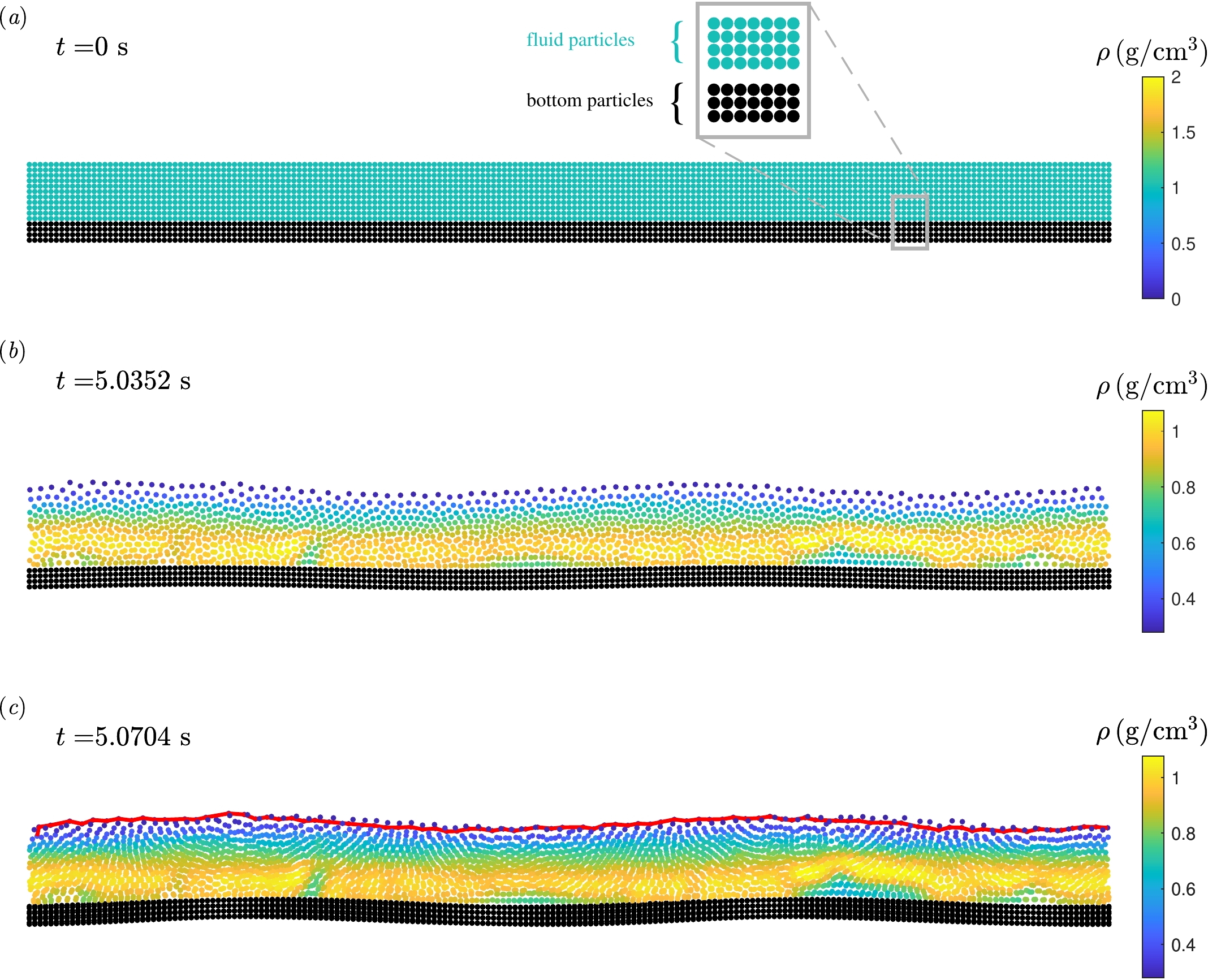}}
    \end{minipage}
    \hfill
    \begin{minipage}{6cm}
    \scalebox{0.37}{\includegraphics{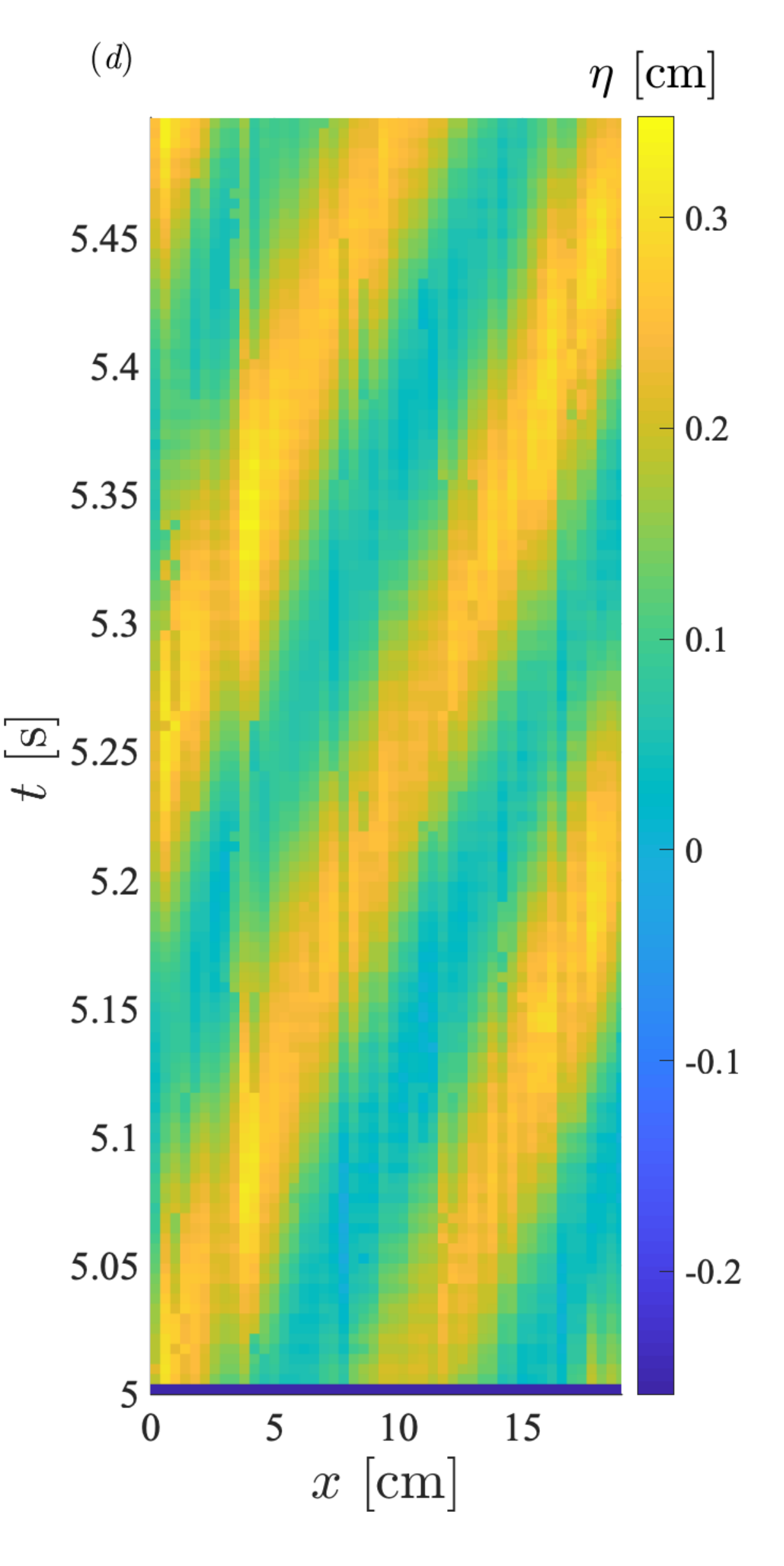}}
    \end{minipage}
    \caption{Numerical protocol for simulating long waves in water channels. \textbf{(\textit{a})} Numerical setup used in SPH simulations. Waves are generated by the pedal wavemakers (black solid particles) at the bottom of a two-dimensional fluid layer (turquoise fluid particles) of depth $\bar{h}=1$ and length $\bar{L}=20$. \textbf{(\textit{b})} The numerical simulation begins with a gradual increase in the amplitude of the bottom motion. At some $t>0$, one observes a regular travelling wave at the free surface \textbf{(\textit{c})} Surface detection algorithm performed on a fully developed gravity wave with $\bar{\lambda}=10$. \textbf{(\textit{d})} Spatiotemporal diagram obtained from the concatenation of consecutively detected free surfaces. The depicted case correspond to $\bar{\omega}=1.01$.}
    \label{Fig:04}
\end{figure*}

We study in SPH simulations the system response in a frequency span of the pedal wavemakers, with $\bar\omega\in[0.1,\,10]$ and $\bar\lambda=10$. Hereon, we shall consider only dimensionless variables unless otherwise stated. At the start of the numerical simulation, the amplitude of the pedal wavemakers gradually increases from zero to a fixed maximum value, $\bar{A}=A/h=0.042$. Thus, we observe the gradual growth of a smooth surface wave without neither defects nor fronts travelling throughout the numerical domain, as depicted in figure~\ref{Fig:04}(\textit{b}). Once a gravity wave is clearly developed, the surface profile is first extracted from a boundary-detection algorithm applied to the set of SPH fluid particles [see figure~\ref{Fig:04}(\textit{c})]. The boundary detection is performed at time steps given by $\Delta\bar{T}=\bar{T}/8$, where $\bar{T}$ is the period of the pedal wavemakers. We made our measurements account for an integer number of oscillations of the pedal wavemakers. Due to particle discreteness, the typical output of this boundary-detection algorithm is an irregular envelope that requires further post-processing. First, we generate a spatiotemporal diagram collecting all the detected boundaries in time, as shown in figure~\ref{Fig:04}(\textit{d}). Then, we compute the two-dimensional Fast Fourier Transform in space and time of such spatiotemporal diagram and find its maximum, which we denote as $\mathcal{F}_{\max}(\eta_{envl})$. Following \eqref{eq:input}, the bottom moves as $\eta_{b}(x,t)=A\cos(kx-\omega t)\propto \mbox{Re}{\,[y_b(x,t)]}$, i.e., a travelling wave forcing towards the positive $x$ direction. We compute similarly the maximum of the fast Fourier transform of this pre-programmed spatiotemporal dynamics corresponding to the pedal wavemakers, $\mathcal{F}_{\max}(\eta_{b})$. Finally, we obtain the gain and the phase of the system with $G=\left\vert \mathcal{F}_{\max}(\eta_{envl})\right\vert/\left\vert\mathcal{F}_{\max}(\eta_{b}) \right\vert$ and $\phi=\arg\left[\mathcal{F}_{\max}(\eta_{envl})/\mathcal{F}_{\max}(\eta_{b})\right]+\Delta\phi_0$, respectively, where $\Delta\phi_0$ is the accumulated phase shift from the beginning of the simulation to the first boundary detection. 

\begin{figure}
    \centering
    \scalebox{0.58}{\includegraphics{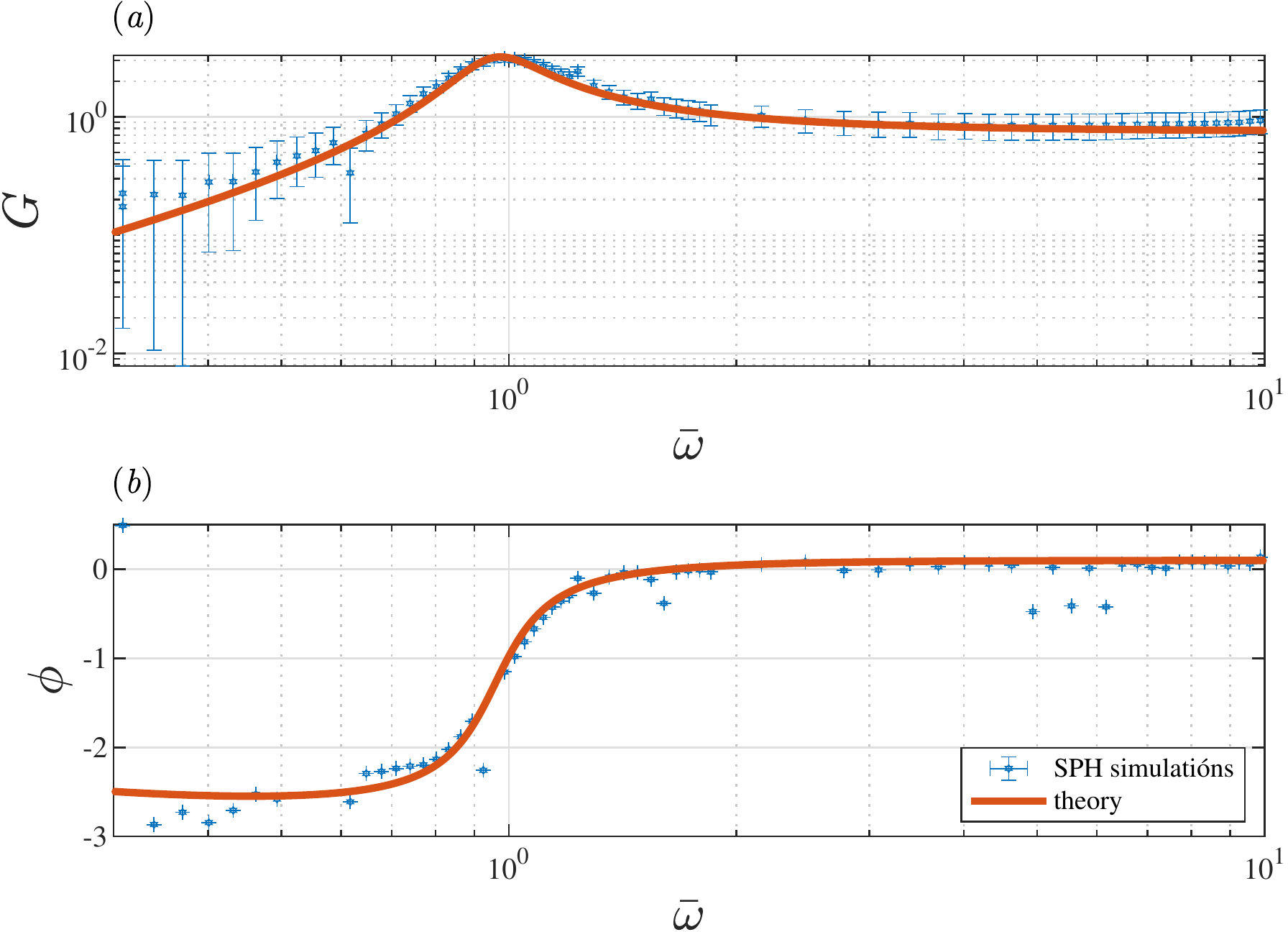}}
    \caption{Gain and phase of the system for $\bar{\lambda}=10$. Comparison between SPH simulations (blue stars) and analytical results using \eqref{Eq:Theoretical} (solid red line). Theoretical curves corresponds to $\bar{\Theta}=0.0958$ ($\nu=1.5\,\mbox{cSt}$).}
    \label{Fig:05}
\end{figure}

Figure \ref{Fig:05} summarises the results from SPH simulations. In figure~\ref{Fig:05}(\textit{a}), we confirm that the system exhibits typical resonance features, where the gain of the system reaches the value $G_{\max}=3.1143$ at $\bar{\omega}=\bar{\omega}_{\hbox{res}}$. Moreover, SPH simulations confirm the high-pass filtering behaviour of the system: the gain strongly decays for decreasing values of $\bar{\omega}$ below $\bar{\omega}_{\hbox{res}}$. In contrast, if the frequency is increased above the resonance, the gain asymptotically decays towards unity. Given that SPH simulations are highly viscous, the resonant peak is relatively broad and the resonant frequency is slightly below the value corresponding to the inviscid case, $\bar{\omega}=1$. Such behaviour was previously observed in figure~\ref{Fig:Resonance_Viscosity}(\textit{f}) in our theoretical study in \S \ref{Sec:Resonance}. In figure~\ref{Fig:05}(\textit{a}) we show in solid red line the predicted curve for the gain corresponding to $\bar{\Theta}=0.0958$, which is in agreement with our theoretical results.

Finally, figure~\ref{Fig:05}(\textit{b}) shows the phase as a function of frequency obtained from SPH simulations, along with the corresponding theoretical curve for the same value of the viscosity, i.e. correspondingly $\bar{\Theta}=0.0958$. We introduced a shift $\Delta\phi=0.21$ in the theoretical curve of the phase to consider numerical effects due to the artificially introduced compressibility of fluid particles in our SPH formulation \citep{Hughes2010}. We observe that the resonance condition occurs slightly below $\bar{\omega}=1$, which is consistent with our previous observations. As expected, SPH simulations confirm that the system synchronises at large frequencies.

\section{\label{Sec:Conclusions}Conclusions}

We studied and characterised the resonance of long gravity graves generated at the free surface of a viscous fluid using pedal wavemakers by analysing linearised Navier-Stokes equations in an infinite two-dimensional fluid layer of finite depth. We calculated a closed-form expression for the system gain as a function of fluid and pedal wavemaking parameters. We obtained that the system exhibits a typical resonance peak and behaves like a high-pass filter in frequency. The resulting resonant peak can be very sharp for small viscosity and display remarkable amplification factors. We showed that pedal wavemakers generate long waves efficiently, e.g. they can generate waves with an amplitude of the order of meters using orbital motions at the bottom with an amplitude of the order of centimetres under ordinary conditions. Moreover, the frequency and wavelength of the generated gravity waves can be tuned to match the dispersion relation of deep-water waves. Fluid parameters affect the resonance: Increasing viscosity value decreases the system gain, widening the resonance peak and introducing a slight shift in the resonant frequency. Accordingly, we have also shown an optimal wavelength for a given combination of viscosity and frequency at which the gain attains the maximum possible value. Thus, the pedal wavemakers are efficient and can mimic the dynamics of waves under deep-water conditions accurately. 

Since our results could be used to design deep-water gravity waves with an on-demand wavelength in a narrow water channel, we put under test pedal wavemakers by performing numerical simulations of the system using the SPH method -a highly viscous particle method. We performed a complete numerical characterisation of the resonant peak, showing excellent agreement with theoretical results even at large viscosity. The results of this work suggest promising applications in the study of oceanic phenomena involving long gravity waves at the surface of the sea, such as ordinary tides, trans-tidal waves, and other waves generated by storms and earthquakes. In addition, our results may shed light on the open question of how relatively small-amplitude perturbations generate large-amplitude surface wave events, such as super-tsunamis appearing after small-intensity earthquakes and rogue waves appearing without warning in otherwise benign conditions.

\begin{acknowledgments}
J.F.M. and A.E. acknowledge the financial support of the National Agency of Research and Development, ANID, through the grant FONDECYT/POSTDOCTORADO/3200499. B.C. thanks to the financial support of the Australian Research Council Linkage Project Number LP190101283. L.G. and I.V. thank the Fondecyt/Iniciación grant No. 11170700.
\end{acknowledgments}


%

\end{document}